\documentclass[11pt]{article}

\usepackage{epsfig}

\usepackage{amsfonts}
\usepackage{amssymb}
\usepackage{amsmath}
\usepackage{epsfig}
\usepackage{color}
\usepackage{graphicx}

\newtheorem{theorem}{Theorem}[section]

\newtheorem{lemma}[theorem]{Lemma}
\newtheorem{claim}[theorem]{Claim}
\newtheorem{fact}[theorem]{Fact}

\newtheorem{conjecture}[theorem]{Conjecture}

\newcommand{\ignore}[1]{ }

\newcommand{\sq}{\hbox{\rlap{$\sqcap$}$\sqcup$}}
\newcommand{\qed}{\hspace*{\fill}\sq}
\newenvironment{proof}{\noindent {\bf Proof.}\ }{\qed\par\vskip 4mm\par}
\newenvironment{proofof}[1]{\bigskip \noindent {\bf Proof of #1:}\quad }
{\qed\par\vskip 4mm\par}

\def\rep{\mbox{rep}}

\def\RR{\mathbb{R}}
\def\ZZ{\mathbb{Z}}

\def\NN{\mbox{NN}}

\def\pr{\mbox{P}}

\def\tiles{\mbox{tiles}}
\def\latt{\mbox{latt}}

\begin{document}

\title{On the metric distortion of nearest-neighbour graphs on random point sets} 

\author{
 Amitabha Bagchi\thanks{Dept of Computer Science and Engg, Indian Institute of Technology, Hauz Khas, New Delhi 110016. {\tt \{bagchi,cs503022\}@cse.iitd.ernet.in.} Note: This work has been submitted to the IEEE for possible
publication. Copyright may be transferred without notice, after
which this version may no longer be accessible.
}
\and Sohit Bansal\footnotemark[1] 
}
\maketitle 

%=========================================================================
%  Abstract
%=========================================================================

\begin{abstract}
We study the graph constructed on a Poisson point process in $d$
dimensions by connecting each point to the $k$ points nearest to
it. This graph a.s. has an infinite cluster if $k > k_c(d)$ where
$k_c(d)$, known as the critical value, depends only on the dimension
$d$. This paper presents an improved upper bound of 188 on the value
of $k_c(2)$. We also show that if $k \geq 188$ the infinite cluster of
$\NN(2,k)$ has an infinite subset of points with the property that the
distance along the edges of the graphs between these points is at most
a constant multiplicative factor larger than their Euclidean
distance. Finally we discuss in detail the relevance of our results to
the study of multi-hop wireless sensor networks.
\end{abstract}

%=========================================================================
%  Introduction
%=========================================================================

\section{Introduction} 
\label{sec:intro}

The $k$-nearest neighbour graph of a point set $S$ in a metric space
is constructed according to the following natural definition: For each
point $x \in S$ establish an edge from $x$ to the $k$ points of $S
\setminus \{x\}$ nearest to it. Such graphs have applications in
numerous areas: classification problems of all flavours, topology
control in wireless networks~\cite{blough-mobihoc:2003,xue-wn:2004},
data compression~\cite{ouyang-wise:2002,adler-dcc:2001} and
dimensionality reduction~\cite{tenenbaum-science:2000} and multi-agent
systems~\cite{goebels-icnc:2006}.

We focus on $k$-nearest neighbor graphs on random point sets in
$\RR^d$ assuming that the distance is the Euclidean distance. Further
we restrict ourselves to the case where the edges established are
undirected. Clearly it is not necessary that this graph be connected
for arbitrary $k$ and $S$ or even that it have a large connected
component. However, H\"aggstr\"om and
Meester~\cite{haggstrom-rsa:1996} have shown that if the set $S$ is
generated by a Poisson point process then there is a finite value
$k_c(d)$ depending only on the dimension such that if $k > k_c(d)$,
the $k$-nearest neighbor graph has an connected component which is
infinite. In this paper we study this setting further. Following the
notation in~\cite{haggstrom-rsa:1996} we will denote this model in $d$
dimensions, parametrized by $k$ as $\NN(d,k)$.

In this paper we show that for $\NN(2,k)$ that if $k \geq 188$ the
infinite cluster\footnote{We will use the terms component and cluster
interchangeably.} has an infinite subset of points with the property
the metric distortion between them is bounded by a constant i.e. if
there is a pair of points in this infinite subset the shortest
distance between them achieved along a path in the graph is at most a
constant multiplicative factor larger than the Euclidean distance
between them. In the process of proving the latter result we improve
the best known bound for $k_c(2)$ to 188 from 213 (due to Teng and
Yao~\cite{teng-algorithmica:2007}).  Our proof technique generalizes
easily to $\NN(d,k)$ for $d \geq 3$.  

{\bf Organization.} The rest of this section is devoted to surveying
related work and introducing the terms and notation we will use. A new
bound on $k_c(2)$ and the result on the metric distortion within the
infinite cluster is presented in Section~\ref{sec:metric}. We will
talk about the applicability of our results to wireless multi-hop
sensor networks in Section~\ref{sec:wireless}, concluding with a
discussion of some simulation results and conjectures arising from
them in Section~\ref{sec:conclusion}.

\subsection{Related work}
\label{sec:intro:related}

The study of random graphs obtained by applying connection rules on
stationary point processes is known as continuum percolation. Meester
and Roy's monograph on the subject provides an excellent view of the
deep theory that has been developed around this general
setting~\cite{meester:1996}. The $\NN(d,k)$ model was introduced by
H\"aggstr\"om and Meester~\cite{haggstrom-rsa:1996}. They showed that
there was a finite critical value, $k_c(d)$ for all $d \geq 2$ such
that an infinite cluster exists in this model. They proved that the
infinite cluster was unique and that there was a value $d_0$ such that
$k_c(d) = 2$ for all $d > d_0$. Teng and Yao gave an upper bound of
213 for $k_c(d)$~\cite{teng-algorithmica:2007}.

$k$-nearest neighbor graphs on random point sets contained inside a
finite region have been extensively studied. The major concern,
different from ours, has been to ensure that {\em all} the points
within the region are connected within the same cluster. Ballister,
Bollob\'as, Sarkar and Walters~\cite{ballister-aap:2005} showed that
the smallest value of $k$ that will ensure connectivity lies between
$0.3043 \log n $ and $0.5139 \log n$, improving earlier results of Xue
and Kumar~\cite{xue-wn:2004}. Ballister et. al. also studied the
problem of covering the region with the discs containing the
$k$-nearest neighbours of the points. We refer the reader
to~\cite{ballister-aap:2005} for an interesting discussion relating
this setting to earlier work by Penrose and others.

Eppstein, Paterson and Yao~\cite{eppstein-dcg:1997} studied
$k$-nearest neighbour graphs on random point sets in two dimensions in
some detail and proved interesting bounds showing that the number of
points in a component of depth $D$ was polynomial in $D$ when $k$ was
1 and exponential in $D$ when it was 2 or greater. Their primary
interest was in obtaining low dilation embeddings of nearest-neighbor
graphs.

Algorithms for searching for nearest neighbors (see
e.g.~\cite{clarkson:2005,vaidya-dcg:1989}) and constructing nearest
neighbor graphs efficiently have also received a lot of attention (see
e.g.~\cite{paredes-wea:2006}). However these are not directly related
so we do not survey this literature in detail.

\subsection{Definitions and Notation}
\label{sec:intro:definitions}

\noindent{\bf Poisson point processes.} Our random point sets are
generated by homogenous Poisson point processes of intensity $\lambda$
in $\RR^d$ where $d\geq1$. Under this model the number of points in a
region is a random variable that depends only on its $d$-dimensional
volume i.e. the number of points in a bounded, measurable set $A$ is
Poisson distributed with mean $\lambda V(A)$ where $V(A)$ is the
$d$-dimensional volume of A. Further, the random variables associated
with the number of points in disjoint sets are independent.

\noindent{\bf Site percolation.} Consider an infinite graph defined on
the vertex set $\ZZ^d$ with edges between points $x$ and $y$ such that
$\| x - y\|_1 = 1$. Site percolation is a probabilistic process
on this graph. Each point of $\ZZ^d$ is taken to be {\em open} with
probability $p$ and {\em closed} with probability $1 - p$. The product
of all the measures for individual points forms a measure for the
space of possible configurations. An edge between two open vertices is
considered open. All other edges are considered closed. A component in
which open vertices are connected through paths of open edges is known
as an open cluster. It is known that there is a value $p_c$ such that
for all $p > p_c$ the graph obtained has an infinite open
cluster. This value is known as the critical probability. When $p >
p_c$ then each point of $\ZZ^d$ has some non-zero probability of being
part of an infinite cluster. The reader is referred
to~\cite{grimmett:1999} for a full treatment of percolation and to
\cite{bollobas:2006} for a recent update on some new directions in
this area.

%=========================================================================
%  A subset of the point set has constant metric distortion
%=========================================================================

\section{An infinite subset of $C_\infty$ has constant metric
  distortion} 
\label{sec:metric}

The graph distance between pairs of points in a $k$-nearest neighbor
graph is clearly at least the Euclidean distance between them. The
question arises if the distance is arbitrarily larger than the
Euclidean. Clearly, for points in different clusters the distance this
question makes no sense. We also ignore for now the question of what
happens inside finite clusters, focussing for now on the infinite
cluster in the supercritical phase of $\NN(d,k)$. We conjecture that
it is possible to show that the distance between any pair of points in
the infinite cluster is only a small factor larger than the Euclidean
distance between them. In this paper we prove a weaker result: the
infinite cluster contains an infinite subset of points whose pairwise
distances are not distorted by more than a constant factor. In order
to do this we first present a construction that allows us to couple
$\NN(2,k)$ with a site percolation process in $\ZZ^2$. This
construction also improves the best known upper bound for
$k_c(2)$. Then we show how to use the algorithm of Angel
et. al.~\cite{angel-podc:2005} for routing on a percolated mesh to
find a short path between a pair of vertices in $\NN(d,k)$.

\subsection{Coupling $\NN(2,k)$ to site percolation in $\ZZ^2$}
\label{sec:metric:2d-bounds}

Like H\"aggstr\"om and Meester's proof for the existence of a critical
value~\cite{haggstrom-rsa:1996} and Teng and Yao's proof for the
weaker of their two upper bounds on
$k_c(2)$~\cite{teng-algorithmica:2007}, we proceed by constructing a
coupling with a site percolation process on $\ZZ^2$. However, our
construction gives a better upper bound than Teng and Yao's
improvement of their own result (also
in~\cite{teng-algorithmica:2007}) to $k_c(2) \geq 213$ which uses a
coupling to a mixed percolation process. We are able to improve this
result to show $k_c(2) \geq 188$. Note that both papers, the one by
H\"aggstr\"om and Meester and the one by Teng and Yao, reported that
simulations seemed to indicate that the value of $k_c(2)$ appears to
be around 3. Our simulations backed up this finding. Let us now
proceed to a formal statement of the main theorem of this section and
it's proof.

\begin{theorem}
\label{thm:2d-bound}
For the $k$-nearest neighbour model in a Poisson point process setting
\[k_c(2) \leq 188.\]
\end{theorem}

\begin{figure}[htbp]
\begin{center}
\input{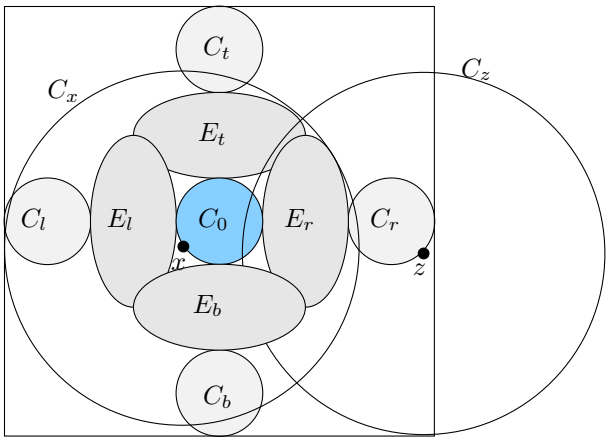}
\caption{A tile $t$ and it's 9 relevant regions. Note that the region
  $E_r$ lies wholly within all discs of the form $C_x$ and $C_z$
  centred at points on the boundary of the discs $C_0$ and $C_r$.}
\label{fig:2d-bound}
\end{center}
\end{figure}

\begin{proof} In order to prove the theorem we couple a site
percolation process on $\ZZ^2$ with the $k$-nearest neighbour graph as
follows. We divide $\RR^2$ into square tiles of side 10$a$ where $a$
is a parameter whose value will be fixed later. We create a bijection,
$\phi$, between these tiles in $\RR^2$ and points in $\ZZ^2$ such that
neighbouring tiles in $\RR^2$ correspond to neighbouring points in
$\ZZ^2$. We couple the processes by saying that a given point $x$ in
$\ZZ^2$ is open only if the tile $t = \phi^{-1}(x)$ a certain event
$A_t$ occurs. We now define this event $A_t$.

Let us look at a tile centred at $(0,0)$ with bottom left corner
$(-5a,-5a)$ and top right corner $(5a,5a)$. For convenience we will
refer to the tiles surrounding the tile $t$ as, couunterclockwise
starting from the right $t_r$, $t_t$, $t_l$ and $t_b$. We consider
five circles of radius $a$: $C_0$ centred at $(0,0)$, $C_l$ centred at
$(-4a,0)$, $C_r$ centred at $(4a,0)$, $C_t$ centred at $(0,4a)$ and
$C_b$ centred at $(0,-4a)$. There are four other region which are
named $E_l, E_r, E_t$ and $E_b$ in the figure. $E_r$ is defined as
follows. Consider the largest circle centred at any point in $C_0$ or
$C_r$ that lies wholly within the two tiles $t$ and $t_r$. Two such
circles, $C_x$ and $C_z$, are depicted in
Figure~\ref{fig:2d-bound}. $E_r$ is the locus of the points contained
in all such circles. The regions $E_l, E_t$ and $E_b$ are defined
similary by $C_0$ alongwith $C_l, C_t$ and $C_b$ respectively and the
tiles $t_l, t_t$ and $t_b$ respectively. 

Now, for a tile $t$, the event $A_t$ is said to occur if
\begin{enumerate}
\item the number of points inside $t$ is at most $k/2$ and 
\item the nine regions $C_0, C_r, C_t, C_l, C_b, E_r, E_t, E_l$ and
  $E_b$ contain at least one point each.
\end{enumerate}

If $A_t$ occurs we call the point contained in $C_0$ the {\em
representative point} of the tile $t$, denoted $\rep(t)$. In order to
relate the process on $\ZZ^2$ defined via these events $A_t$ to the
$\NN(d,k)$ model, we claim that the existence of an edge in $\ZZ^2$
implies the existence of a path from the representative points of the
two tiles corresponding to the two end points of the edge. We state
this formally, including an observation about the metric distortion of
the length of the path between the two representative points.

\begin{claim}
\label{clm:coupling}
If an edge exists in the percolated mesh $\ZZ^2$ between two points
$x$ and $y$  then
\begin{enumerate}
\item There is a path between the representative points
  $\rep(\phi^{-1}(x))$ and $\rep(\phi^{-1}(y))$ of the tiles
  corresponding to $x$ and $y$ in $\NN(2,k)$ and
\item there is a constant $c_{\tiles}$ such that 
\[d_k(\rep(\phi^{-1}(x)), \rep(\phi^{-1}(y))) \leq c \cdot
d(\rep(\phi^{-1}(x)), \rep(\phi^{-1}(y))).\] 
\end{enumerate}
\end{claim}

\begin{figure}[htbp]
\begin{center}
\input{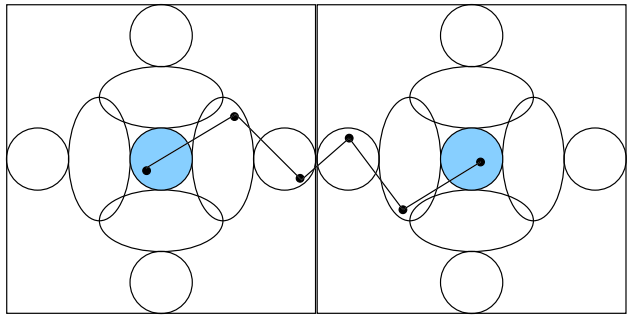}
\caption{A path between two representative points of tiles for both of which
  the event $A_t$ has occured.} 
\label{fig:coupling-claim}
\end{center}
\end{figure}

\begin{proofof}{Claim~\ref{clm:coupling}} The proof of the claim is
  depicted in Figure~\ref{fig:coupling-claim}
Clearly any circle drawn from $\rep(t)$ that stays within $t$ contains
all of $E_r$ in it by the definition of $E_r$. Since there are at most
$k/2$ points in every tile for which $A_t$ has occured, hence there is
an edge from $\rep(t)$ to the point guaranteed to be contained in
$E_r$, let's call it $x_r$, by the definition of $A_t$. We do not make
any claims on where the edges established by $x_r$ to its neighbours
lie, observing only that any point that lies in $C_r$ must have an
edge to $x_r$, again by the definition of $E_r$. However, any disc
centred at a point in $C_r$ that remains within $t$ and $t_r$ must
contain the left disc of its neighboring tile. Hence, if $A_t$ and
$A_{t_r}$ occur then a path from $\rep(t)$ to $\rep(t_r)$ occurs. The
second part of the claim is obviously true. The constant can easily be
calculated using calculus.
\end{proofof}

From Claim~\ref{clm:coupling}, it is easy to deduce that if an
infinite component exists in the site percolation setting, then an
infinte component exists in $\NN(2,k)$. Hence we need to determine for
what settings of our parameters $a$ and, more importantly, $k$, the
site percolation process is supercritical.  The critical probability
for site percolation is 0.59 (see
e.g. \cite{lee-arxiv:2007}). Numerical calculations showed that the
smallest value of $k$ for which the probability of $A_t$ exceeds this
value is 188, and the value of $a$ for which this happens is 0.893.
\end{proof}

\subsection{A subset with constant metric distortion}
\label{sec:metric:subset}

We now show that there is a set of points in $C_\infty$ and constant
$\alpha$ such that for each pair of points $x,y$ in this set
\[ D_k(x,y) \leq \alpha \cdot D(x,y).\]
We will prove the following theorem
\begin{theorem}
\label{thm:metric}
For $\NN(2,k)$ where $k > 188$, there is a set of points $S
\subseteq C_\infty$ such that $|S| = \infty$ with the following
property: Let $x,y \in S$ be two points with Euclidean distance
$D(x,y)$ between them whose $k$-NN distance is $D_k(x,y)$. For some
$\alpha, c$ depending only on $k$
\[\pr(D_k(x,y) > \alpha \cdot D(x,y)) < e^{-c \cdot D(x,y)}.\]
\end{theorem}
\begin{proof}
We identify $S$ to be the set of representative points lying in the
infinite cluster of $\NN(2,k)$ of the construction described in the
proof of Theorem~\ref{thm:2d-bound} as the subset that we will claim
has this property. We use the coupling with site percolation in
$\ZZ^2$ introduced in that proof to help us find short paths between
pairs of points in $S$. 

Let us consider any two tiles $t_1$ and $t_2$ whose representative
points $\rep(t_1)$ and $\rep(t_2)$ lie in $C_\infty$. We denote
distance between two points $a,b$ in $\ZZ^2$ is denoted
$D_{\latt}(a,b)$. First we relate the distance in the (unpercolated)
lattice to the euclidean distance between these two points by
observing a simple fact.
\begin{fact}
\label{fct:euclidean-lattice}
Given that $c_{\tiles}$ is the constant defined in
Claim~\ref{clm:coupling} then for two tiles $t_1, t_2$
\[D_{\latt}(\phi(\rep(t_1)), \phi(\rep(t_2))) \leq \sqrt{2} \cdot
\frac{D(\rep(t_1),\rep(t_2))}{c_{\tiles}}.\]
\end{fact}

When the lattice undergoes percolation, the simple open path from
$\phi(\rep(t_1))$ to $\phi(\rep(t_2))$ may be broken at several
points. Antal and Pisztora studied this setting and proved a powerful
theorem which helps us here~\cite[Theorems 1.1 and
1.2]{antal-ap:1996}. We use it as a lemma here, adopting the
restatement of Angel et. al.~\cite[Lemma 8]{angel-podc:2005}.
\begin{lemma}
\label{lem:antal}
{\bf \cite{antal-ap:1996,angel-podc:2005}} For any $p > p_c$ and any
  $x,y$ connected through an open path in a cube $M^d$ of the infinite
  lattice, let $D^p_{\latt}(x,y)$ be the distance between the two
  points in the percolated lattice. For some $\rho, c_2 > 0$ depending
  only on the dimension and $p$ and for any $a > \rho \cdot
  D_{\latt}(x,y)$
\[pr(D^p_{\latt}(x,y) > a)) < e^{-c_2 a}.\]
\end{lemma}

\begin{figure}[htbp]
\begin{center}
\input{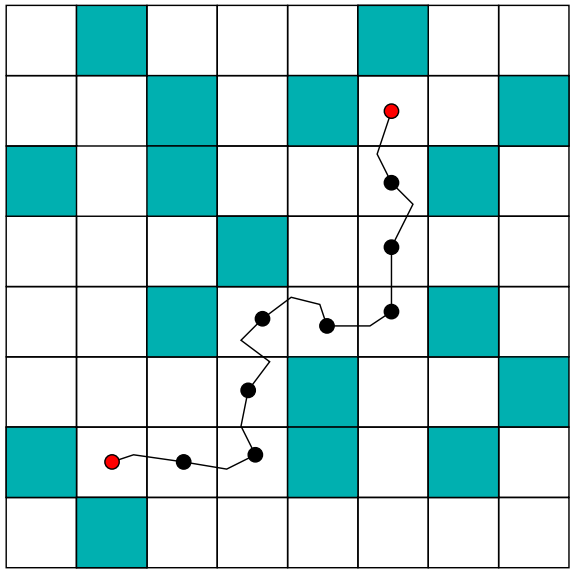}
\caption{The path between two representative points mimics the path in
  $\ZZ^2$.} 
\label{fig:metric-path}
\end{center}
\end{figure}

To find a path between $\rep(t_1)$ and $\rep(t_2)$ we simply take the
path in the percolated lattice between $\phi(t_1)$ and $\phi(t_2)$ and
mimic it $\RR^2$ as depicted in Figure~\ref{fig:metric-path} and the
result follows by combining Fact~\ref{fct:euclidean-lattice} and
Lemma~\ref{lem:antal}.
\end{proof}

We note here that our claim that the constant in the statement of
Theorem~\ref{thm:metric} depends only on the value of $k$ follows from
the fact that the constants in Lemma~\ref{lem:antal} depend only on
$p$, since in our construction the size of a tile and the probability
of $A_t$ occuring for a tile changes when we change $k$. We also note
that Antal and Pisztora~\cite{antal-ap:1996} prove their theorem for
bond percolation but note that their methods can easily be extended to
site percolation.

It is possible to extend Theorem~\ref{thm:metric} easily for $d >
2$. The constants change and their dependence on $d$ has to be handled
carefully but the proof remains basically the same.

%=========================================================================
%  Elaboration of applications to wireless networks
%=========================================================================

\section{Applications to multi-hop wireless sensor networks.} 
\label{sec:wireless}

Multi-hop sensor networks, where nodes act not only to sense but also
to relay information, have proven advantages in terms of energy
efficiency over single hop sensor networks~\cite{karl:2005} and are
useful necessary tasks like time
synchronization~\cite{vangreunen-wsna:2003}. And for collaborative
tasks like target tracking~\cite{zhao-ieee:2003} sensor-to-sensor
communication is essential. But the total connectivity sought to be
achieved in~\cite{xue-wn:2004,ballister-aap:2005} between all the
points of a point process is not necessary for these networks. It may
be the correct model for general ad hoc wireless networks where all
nodes need to be connected, but for a sensor network we argue the
presence of large connected component is enough.

Sensor networks seek to achieve coverage of a target area. When the
locations of sensors are modelled by point processes achieving most
coverage measures (whether it is single point coverage or $k$-coverage
or barrier coverage) has found that there is a critical density of the
point process above which the particular measure is satisfactory. For
example~\cite{ballister-mobicom:2007} estimates the critical density
required for barrier coverage in strip-like regions, a notion of
coverage where an object must be sensed if it tries to cross a
particular region.\footnote{See~\cite[Chap 13.2]{karl:2005} for a
succinct summary of the issues involved in coverage.}

Our results show that it is possible to find an infinite component
with which has an infinite subset of nodes whose graph distance is a
constant times their euclidean distance. Our construction for the
proofs of Theorems~\ref{thm:2d-bound} and \ref{thm:metric}, taken
along with the fact that for any point in $\ZZ^2$ there is a non-zero
probability of being part of the infinite component in the
supercritical phase imply the following theorem

\begin{theorem}
\label{thm:sensor-coverage}
For any $\lambda$, there is a $\lambda'$ such that $\NN(2,k)$ built on
a point process of density $\lambda'$ with $k>188$, has an infinite
component with the property that an infinite subset of points with
density at least $\lambda$ has the property that that graph distance
between them is at most a constant times the Euclidean distance
between them. Moreover there is a constant $c$ such that $\lambda' < c
\lambda$. 
\end{theorem}

Clearly the existence of such a subset can fulfil the sensing
requirement while not compromising on the sensor-to-sensor data
transfer capability.  The value 188 seems prohibitive for most
practical purposes. But it is our hope that this upper bound will be
improved down to a reasonable value closer to the 2 conjectured by
H\"aggstr\"om and Meester~\cite{haggstrom-rsa:1996} and Teng and
Yao~\cite{teng-algorithmica:2007} and that it will be possible to
prove Theorem~\ref{thm:metric} for this improved bound as well. We
omit the proof of this theorem here since it does not add any new
insight over the proofs already seen in this paper.

%=========================================================================
%  Conclusion
%=========================================================================

\section{Conclusion}
\label{sec:conclusion}

We conclude by presenting some conjectures about the relationship of
the metric distortion in $\NN(2,k)$ to the parameter $k$. These
conjectures come from simulations we ran.

The experiments had to be carried out on a finite box in $\RR^2$, but
to negate boundary effects we simulated a point process in a large box
and looked at the largest component formed within a smaller box
contained well within this finite box. We placed a number of points
randomly within the larger area (thereby achieving a target
density). In 
Table~\ref{tab:metric-distortion} the first column has the number of
points placed. The table shows the average distortion for different
values of $k$, maximum value of the distortion and the 
percentage of points having distortion less than two times the
average. This table also indicates that there the distortion is
independent of the number of points under consideration but depends on
the value of $k$.

\begin{table}
\begin{center}
\begin{tabular}{|c|cccc|}
\hline
  $n$& $k$ & avg.& max. value & percentage \\
\hline
  $ 500 $& 3& 1.727& 15.180& $96.96\%$\\
\hline
  $ 500 $& 4& 1.364& 7.543& $97.96\%$\\
\hline
  $ 500 $& 5& 1.204& 5.874& $99.38\%$\\
\hline
  $ 1000 $& 3& 1.660& 22.64& $97.08\%$\\
\hline
  $ 1000 $& 4& 1.333& 8.39& $98.92\%$\\
\hline
  $ 1000 $& 5& 1.172& 4.385& $99.82\%$\\
\hline
  $ 1500 $& 4& 1.322& 7.858& $99.12\%$\\
\hline
  $ 2000 $& 4& 1.285& 9.512& $99.4\%$\\
\hline
\end{tabular}
\end{center}
\label{tab:metric-distortion}
\caption{Metric distortion in $\NN(2,k)$. The last column shows the
  percentage of pairs distorted by a factor of 2 or less.}
\end {table}

\begin{figure}
  \begin{center}
    \includegraphics[scale=0.8]{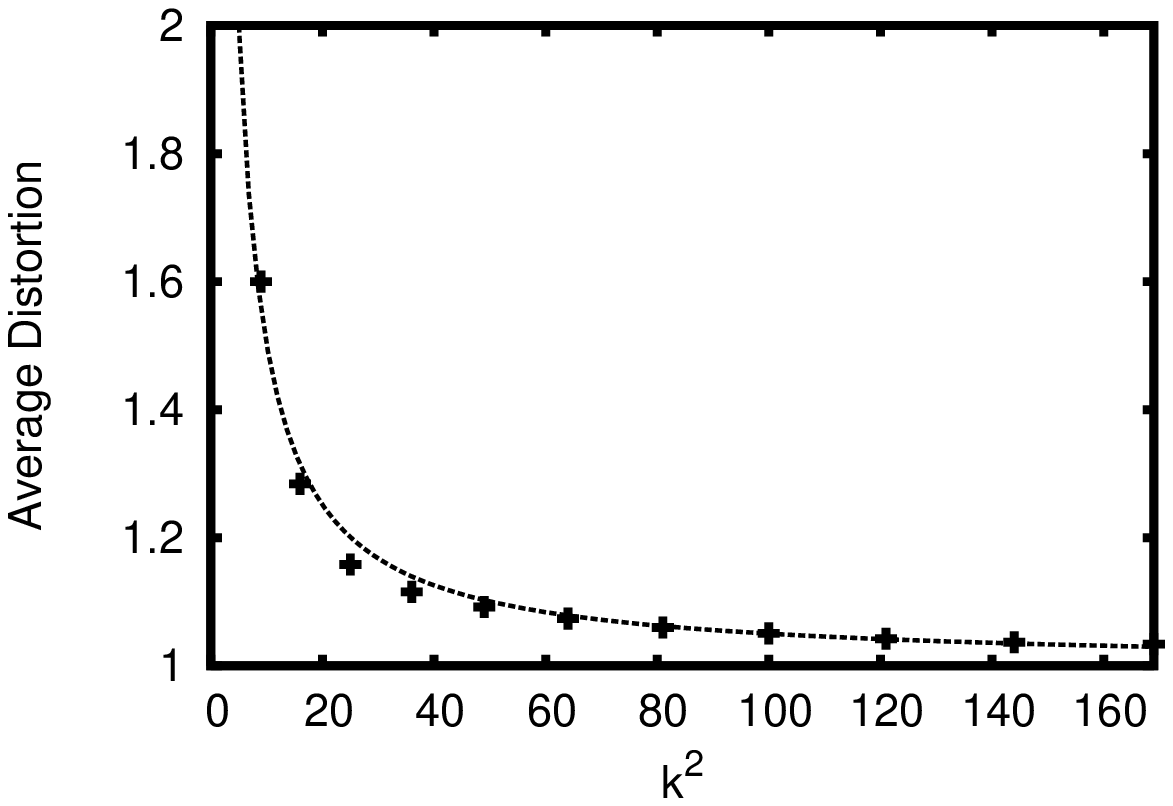}
   \end{center}
\label{fig:plot-k-1}
\caption{Average metric distortion on the $y$-axis and $k^2$ on the
  $x$ axis. The curve plotted is 1 + 4.62/$k^2$.}
\end{figure}

\begin{figure}
  \begin{center}
    \includegraphics[scale=0.8]{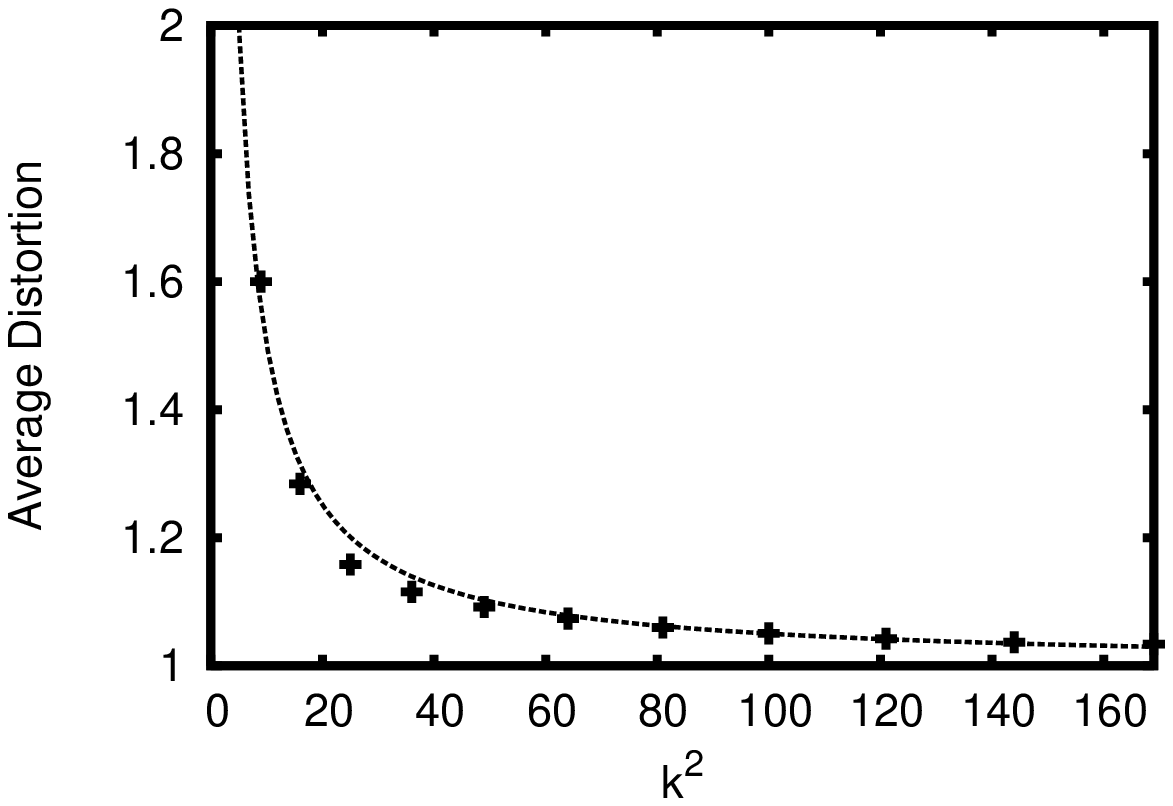}
  \end{center}
\label{fig:plot-k-2}
\caption{Average metric distortion on the $y$-axis and $k^2$ on the
  $x$ axis. The curve plotted is 1 + 5.03/$k^2$.}
\end{figure}

To show relationship between $k$ and average distortion we plotted
average ratio with $k^2$ for a range of value of k from 3 to 13 for
two random point sets. Figures~\ref{fig:plot-k-1}
and~\ref{fig:plot-k-2} show plots for two such sets along with a
function $f(k) = 1 + a/k^2$ where $a$ is determined by least square
fitting functions. These findings lead us to conjecture that:

\begin{conjecture}
For the $\NN(2,k)$ model at a value $k > k_c(2)$
\begin{enumerate}
\item The metric distortion of the points of $C_\infty$ is at most 2
  with probability tending to 1 and 
\item there is a constant such that the expected metric distortion of
  the points of $C_\infty$ is of the form $1 + \frac{a}{k^2}$.
\end{enumerate}
\end{conjecture}

%=========================================================================
%  Bibliography
%=========================================================================

% \newpage
\bibliographystyle{abbrv}

\end{document}